\RequirePackage{lineno}
\pdfoutput=1
\documentclass[number,twocolumn,showpacs,showkeys,amsmath,amssymb,superscriptaddress,groupedaddress,showkeys]{revtex4}  
\usepackage{graphicx}  
\usepackage{epstopdf}
\usepackage{subfigure}
\usepackage{dcolumn}   
\usepackage{bm}        
\usepackage{amssymb}   
\usepackage{lineno}
\usepackage{multirow}
\usepackage{indentfirst}
\usepackage{subfigure}
\usepackage{array}
\usepackage{xcolor}

\newcommand{\D}{\displaystyle}
\newcommand{\DF}[2]{\frac{\D#1}{\D#2}}

\begin{document}
\setlength{\parskip}{0\baselineskip}
\title{Wide field-of-view and high-efficiency light concentrator}

\newcommand{\KeyLab}  {\affiliation{Key Laboratory of Particle \& Radiation Imaging (Tsinghua University), Ministry of Education, Beijing 100084, China}}
\newcommand{\TsingHua}{\affiliation{Department~of~Engineering~Physics, Tsinghua~University, Beijing 100084, China}}
\newcommand{\YZN}{\affiliation{Key Laboratory of Nuclear Data, China Institute of Atomic Energy, Beijing 102413, China}}

\author{Yu~Zhi}\TsingHua\KeyLab\YZN
\author{Ye~Liang}\TsingHua\KeyLab
\author{Zhe~Wang\footnote{Corresponding author: wangzhe-hep@mail.tsinghua.edu.cn}}\TsingHua\KeyLab
\author{Shaomin~Chen}\TsingHua\KeyLab

\date{\today}
\begin{abstract}
  To improve light yield and energy resolution in large-volume neutrino detectors, light concentrators are often mounted on photomultiplier tubes to increase the detection efficiency of optical photons from scintillation or Cherenkov light induced by charged particles.
  We propose a method to optimize previous light concentrators design in order to attain a field of view of 90$^\circ$ and a geometrical collection efficiency above 98\%. This improvement could be crucial to Jinping and other future neutrino experiments whichever it is applicable.
\end{abstract}
\maketitle
\section{Introduction}
Many neutrino detectors use water, heavy water or liquid scintillator as neutrino target and detection material.
Cherenkov and scintillation light induced by charged particles, products of neutrino interaction, are detected by photomultiplier tubes (PMTs).
Light concentrators (or reflectors), developed based on Winston cone~\cite{Welford, Guan:2006gsa}, have been adopted to be mounted on PMTs by several neutrino experiments, for example, the SNO~\cite{SNO, SNOThesis} and Borexino~\cite{BX} experiments, and cosmic ray telescopes~\cite{SPIE, ASP, POS}.
Water-based liquid scintillator or slow liquid scintillator, both of which feature Cherenkov and scintillation separation,
may be available and very interesting in the near future~\cite{Minfang, LAB, GuoZY, CHESS, Hanyu, GeoMK}.
To detect sufficient light and to achieve a high energy resolution for solar neutrino studies using a slow liquid scintillator, the Jinping~\cite{Jinping, JPGeo} neutrino experiment is also considering the use of light concentrators.
Other neutrino experiments are also interested in light concentrators.
It may be the default option in the LENA experiment~\cite{LENA}.
In parallel to this study for Jinping, similar R\&D activities on light concentrators are also being developed for the JUNO experiment~\cite{JUNO, JUNO2}.

\begin{figure}[h]
\centering
\includegraphics[width=0.5\textwidth]{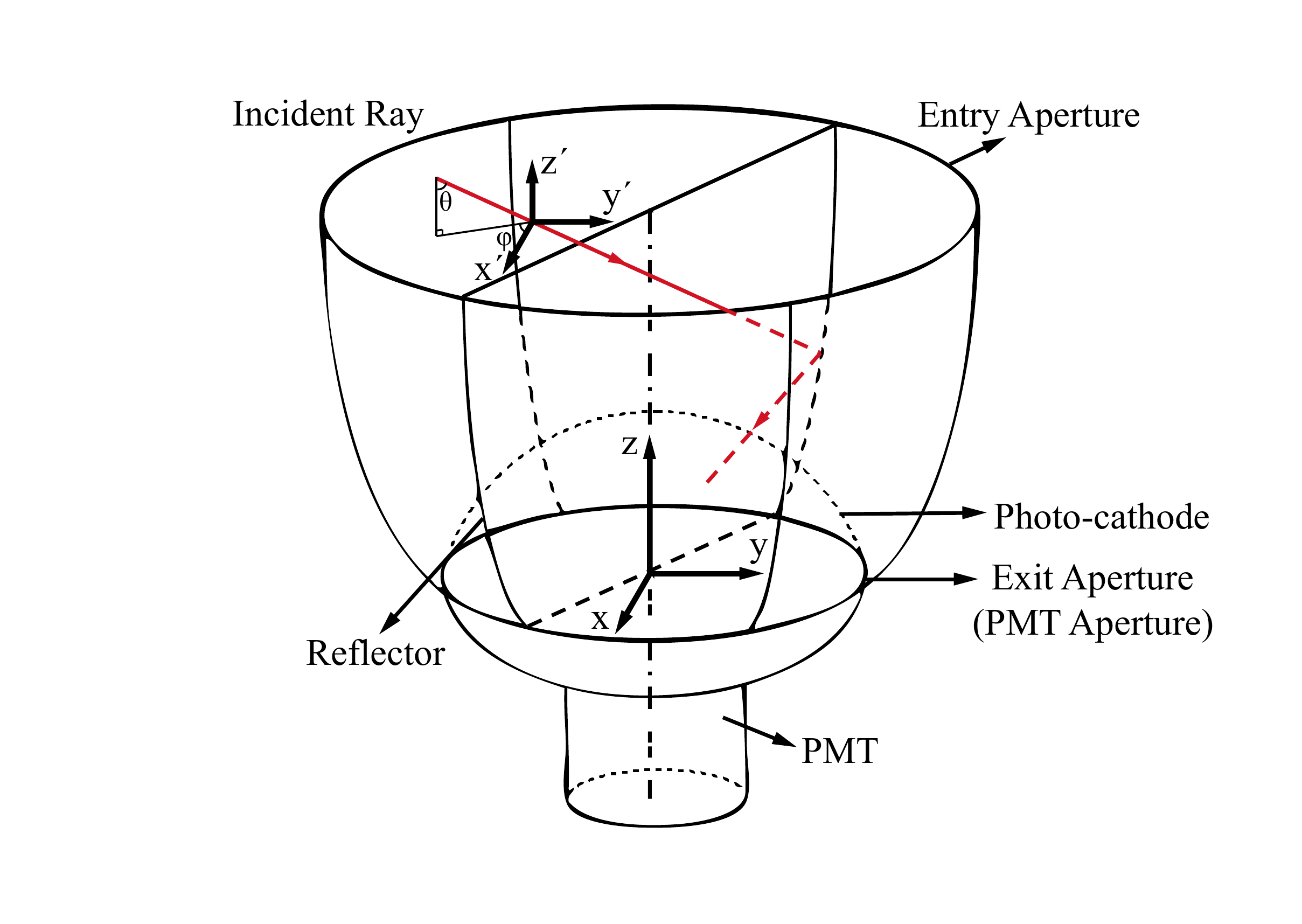}
\caption{Illustration of a PMT with a concentrator. The incident angle, $\theta$, is the angle between the incident ray (red line) and the symmetry axis (z axis).
The azimuth angle, $\phi$, is the angle between the x-axis and the projection of incident ray on the x-y plane.
The x-axis is arbitrary unless specified.
The entry aperture, exit aperture (PMT aperture) are parallel to the x-y plane.
More details are explained in the text.}
\label{fig:intro}
\end{figure}

Figure~\ref{fig:intro} shows how a light concentrator is used with a PMT.
By design, incident light that may have missed the photo cathode can be reflected on to it, if the incident angle $\theta$ (see Fig.~\ref{fig:intro} for the conventional definition of $\theta$ and $\phi$) of the
light is within a cut-off angle, $\theta_{cut-off}$.
In principle, no acceptance of photons occurs beyond the cut-off angle.
The technique effectively enlarges the aperture of a PMT
by a significant factor, for example, 1.8 for SNO, and 2.7 for Borexino.
The use of light concentrators is favored because of the low cost compared with the expense of increasing the number of PMTs or pursuing larger PMT diameters.

The neutrino detectors of future experiments require a high light collection efficiency and a large target mass~\cite{JUNO, Jinping}.
When employing light concentrators in a very large neutrino detector, for example 10 m diameter for a central target region, a wide field of view is needed.
In previous cited experiments, the detector configuration requires a $\theta_{cut-off}$ of about 50$^\circ$~\cite{SNO, SNOThesis, BX, SPIE, ASP, POS}. The used design method, also known as the String method, can be further considered to achieve better performance. In particular, we focus on two aspects:
1) Within the cut-off angle, the perfect light collection efficiency designed in two dimensions (2D) cannot be preserved in three dimensional (3D) condition~\cite{Welford, Guan:2006gsa, SNO}.
This obstacle is especially serious for wide-view concentrators.
2) Light concentrators with circular apertures cannot achieve a gapless configuration.
The hexagonal design in Cherenkov telescope experiments may solve this problem~\cite{SPIE, ASP, POS}.

In Section~\ref{sec:design}, we further explore the defect of the String method,
introduce a modification in its application, and explore the effects of addition of a hexagonal opening.
In Section~\ref{sec:performance}, the performance and cost of different designs are compared.
Finally, discussions and conclusions are presented in Section~\ref{sec:conclusion}.

\section{Design method}
\label{sec:design}

In this section, the detection efficiency of light concentrators is first defined.
The simulation tools for analyzing concentrators are then explained.
The defect of the String method is explained.
Finally a modified method and a hexagonal light concentrator are introduced.

\subsection{Detection efficiency}
The light detection efficiency, $\varepsilon$, of a detector configuration with light concentrators can be expressed as
\begin{equation}
\label{eq:eff}
\begin{split}
\varepsilon &= \DF{N_{PMT} \cdot S_{PMT} \cdot Amp }{S} \cdot \varepsilon_{ref} \cdot \varepsilon_{col}, \\
            &= \DF{N_{PMT} \cdot S_{PMT} \cdot S_{entry}/S_{exit}}{S} \cdot \varepsilon_{ref} \cdot \varepsilon_{col},\\
            &= coverage \cdot \varepsilon_{ref} \cdot \varepsilon_{col},
\end{split}
\end{equation}
where $N_{PMT}$ is the total number of PMTs with concentrators, $S_{PMT}$ is the area of the exit aperture of the concentrator, i.e.~the area of the photo cathode if it is treated as a flat disk (see Fig.~\ref{fig:intro}), $Amp$ gives the ratio of the entry and exit aperture areas of the concentrator (the exit aperture is identical to the PMT aperture), $S$ is the total surface area of the detector, that will be filled with PMTs and concentrators, $\varepsilon_{ref}$ is the reflectivity of the concentrator and $\varepsilon_{col}$ is the geometrical collection efficiency for all the optical photons upon the entry aperture and within the cut-off angle, $\theta_{cut-off}$.

The first term of equation~(\ref{eq:eff}) gives the effective coverage of all photo cathode,
$\varepsilon_{ref}$ is close to 90\% for aluminum coatings, and is not the emphasis of this article,
and the last term gives the geometrical acceptance for photon detection with a single concentrator.
The total light detection efficiency of a detector is proportional to $\varepsilon$.
Without light concentrators, $Amp$, $\varepsilon_{ref}$ and $\varepsilon_{col}$ are all one, and $\varepsilon$ is simply the photo cathode coverage ${N_{PMT} \cdot S_{PMT}}/{S}$.

$N_{PMT}$ typically has a significant impact on the total cost of an experiment, and $Amp$ and $\varepsilon_{col}$ are the two critical properties of a light concentrator to be optimized.

\subsection{Simulation tools and setup}
The concentrator geometry was first designed in SolidWorks, a software program commonly used for solid modeling. The SolidWorks model was then transformed into triangular facets in FASTRAD, which is a 3D CAD tool for radiation shielding analysis. The output data was used to build concentrator geometries in Geant4~\cite{g41, g42} through the G4TessellatedSolid class.
We used Geant4 simulation to analyze each concentrator.
Light rays were generated uniformly on the entry aperture of the light concentrator, and incident angles were set according to the interest of test.

The sensitive part of the photocathode geometry was approximated as a spherical section with a diameter of 28 cm and a height of 10.46 cm; this geometry coincides with that of an XP1807 PMT~\cite{XP1807}.
We set the reflectivity of the concentrator to be one.

\subsection{Introduction to the String method}

The String method is an improved design method of concentrators based on the Compound Parabolic Curve (CPC, also known as the Winston Cone~\cite{Welford,Guan:2006gsa}) method. Compared with the CPC method, the String method considers the shape of the PMT photo cathode, which results in an increase in the area of the entry aperture and allows a reduction of the number of PMTs per unit area.

 \begin{figure}[h]
 \centering

 \includegraphics[width=7.5cm]{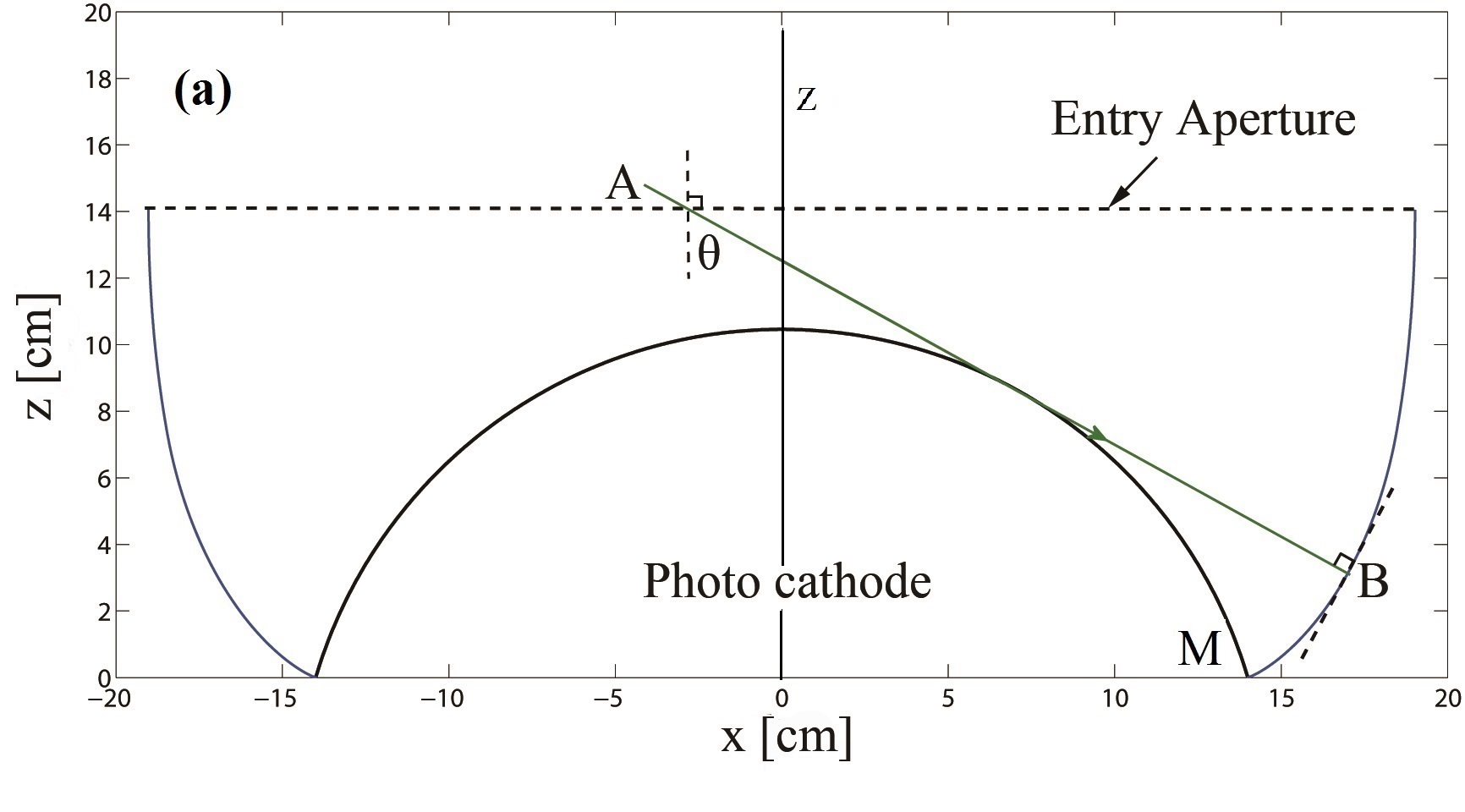}

 \includegraphics[width=7.5cm]{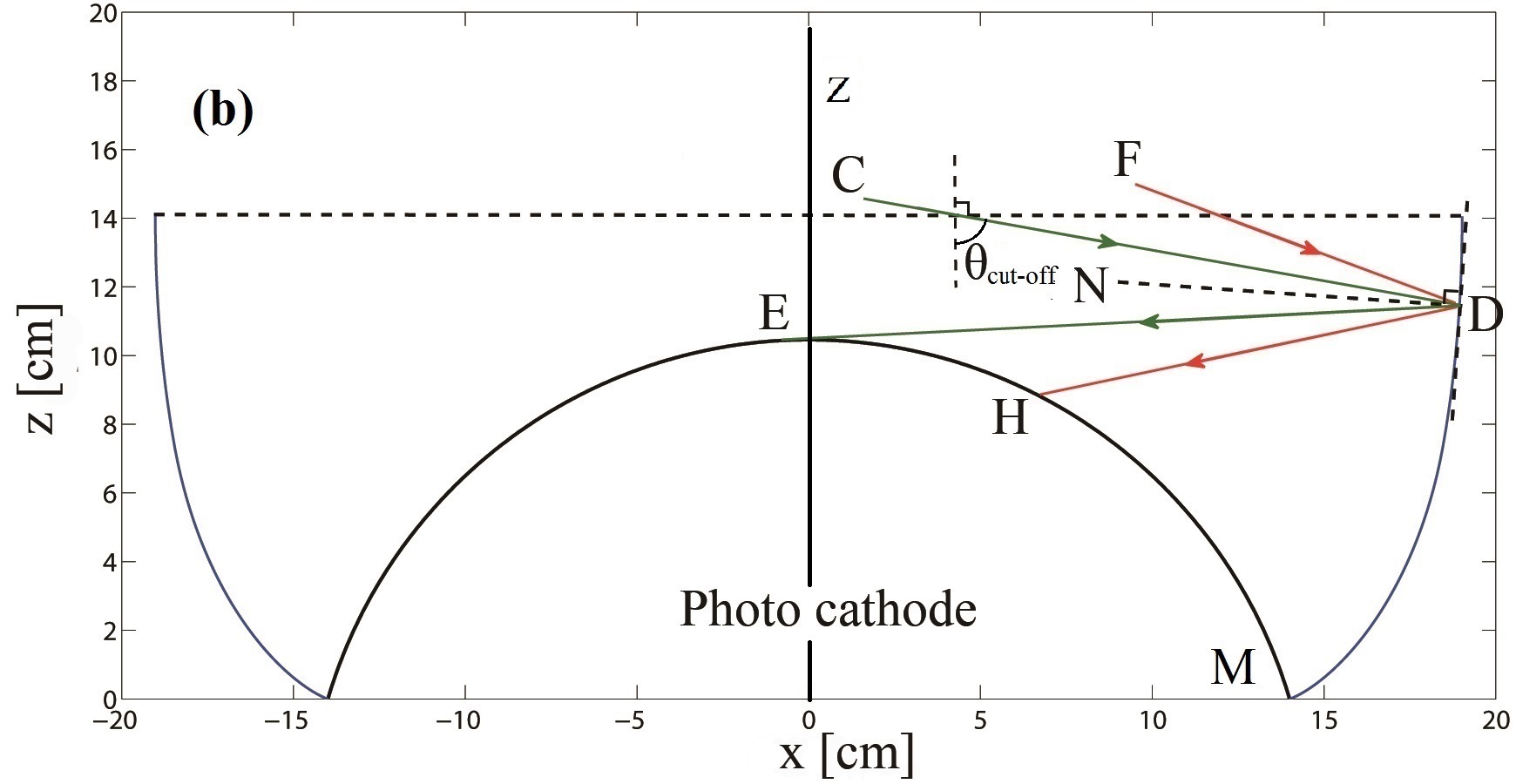}
 \caption{Illustration of the principle of the String method.
 In panel (a) AB is the critical ray for point B, where the incident angle $\theta$ $<$ $\theta_{cut-off}$.
 The slope of the concentrator at B is perpendicular to AB.
 In panel (b) CD is a incident ray, where $\theta = \theta_{cut-off}$.
 ED is the reflection of CD and the critical ray. ND is the angular bisector of CD and ED.
 The slope of the concentrator is perpendicular to ND.
 Thus any rays with $\theta < \theta_{cut-off}$ can reach the photo cathode, similar to FD's reflection onto HD.
 See the text for more details.}
 \label{fig:edgeray}
 \end{figure}

In the String method, given $\theta_{cut-off}$,
the reflector surface is obtained by rotating a 2D reflector profile (Fig.~\ref{fig:edgeray}) around its symmetrical z axis.
The way to construct the profile curve is explained below.

For any point on the reflector curve, a critical ray is a ray starting from the point and tangential to the photo cathode.

Starting with point M (see Fig.~\ref{fig:edgeray}, a), if the incident angle $\theta$ of the critical ray is less than $\theta_{cut-off}$, the slope of the reflector curve must be perpendicular to the critical ray.
An infinitesimal step is added along the slope just determined to create the next point of the curve.
Such a process continues until the incident angle $\theta$ of the critical ray equals $\theta_{cut-off}$.
Afterward the slope must be perpendicular to the angular bisector between the critical ray and
the ray with $\theta=\theta_{cut-off}$ (see Fig.~\ref{fig:edgeray}, b).
The iteration continues until the curve is perpendicular to the entry aperture plane.
Any incident rays with $\theta<\theta_{cut-off}$ will directly arrive at or be reflected onto the photo cathode in the 2D plane.

\begin{figure}[!h]
 \centering
     \includegraphics[width=7cm]{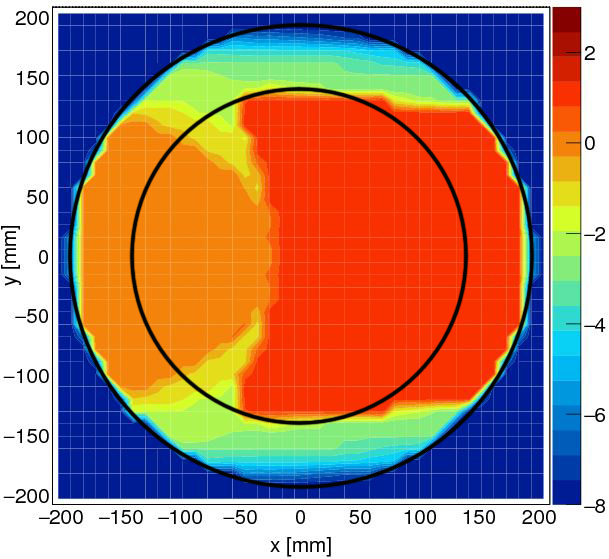}\\
     \includegraphics[width=7cm]{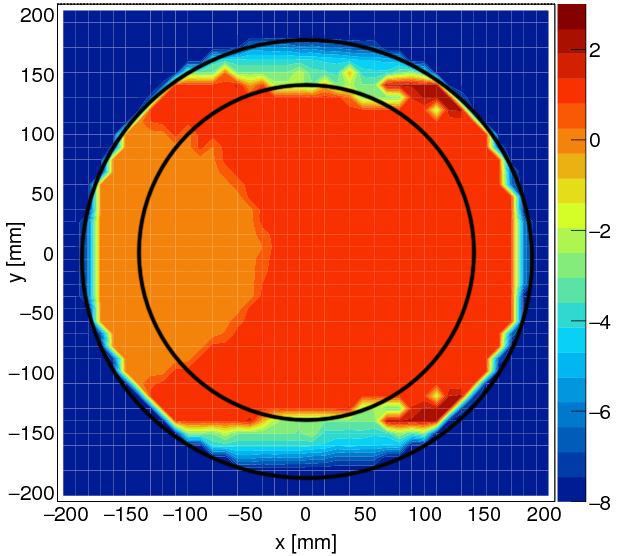}
 \caption{
 Efficiency study for an 80$^\circ$ cut-off angle concentrator.
 All photons feature identical incident angles of 60$^\circ$ and azimuthal angles of 0$^\circ$.
 Each photon incident upon the entry aperture of the String (top) and modified (bottom) designs was traced.
 The color code indicates the number of reflections, and non-negative values indicate that the photon arrives at the photo cathode.
 Entry and exit apertures are presented as black circles.
}
 \label{fig:NoReflection}
 \end{figure}

In the 3D case, this simple feature is not preserved. For example in Fig.~\ref{fig:NoReflection},
for a concentrator with $\theta_{cut-off}=80^\circ$, all photons incident upon the entry aperture with $\theta=60^\circ$ and $\phi=0^\circ$
are traced. Photons with small or large y values may miss the PMT photo cathode,
because the arch of the photo cathode off the central z axis is not as high as that in the 2D profile design. The 3D geometry of the photo cathode will be considered next.

\subsection{3D Modification of the String method}

\begin{figure}[h]
\includegraphics[width=7.5cm]{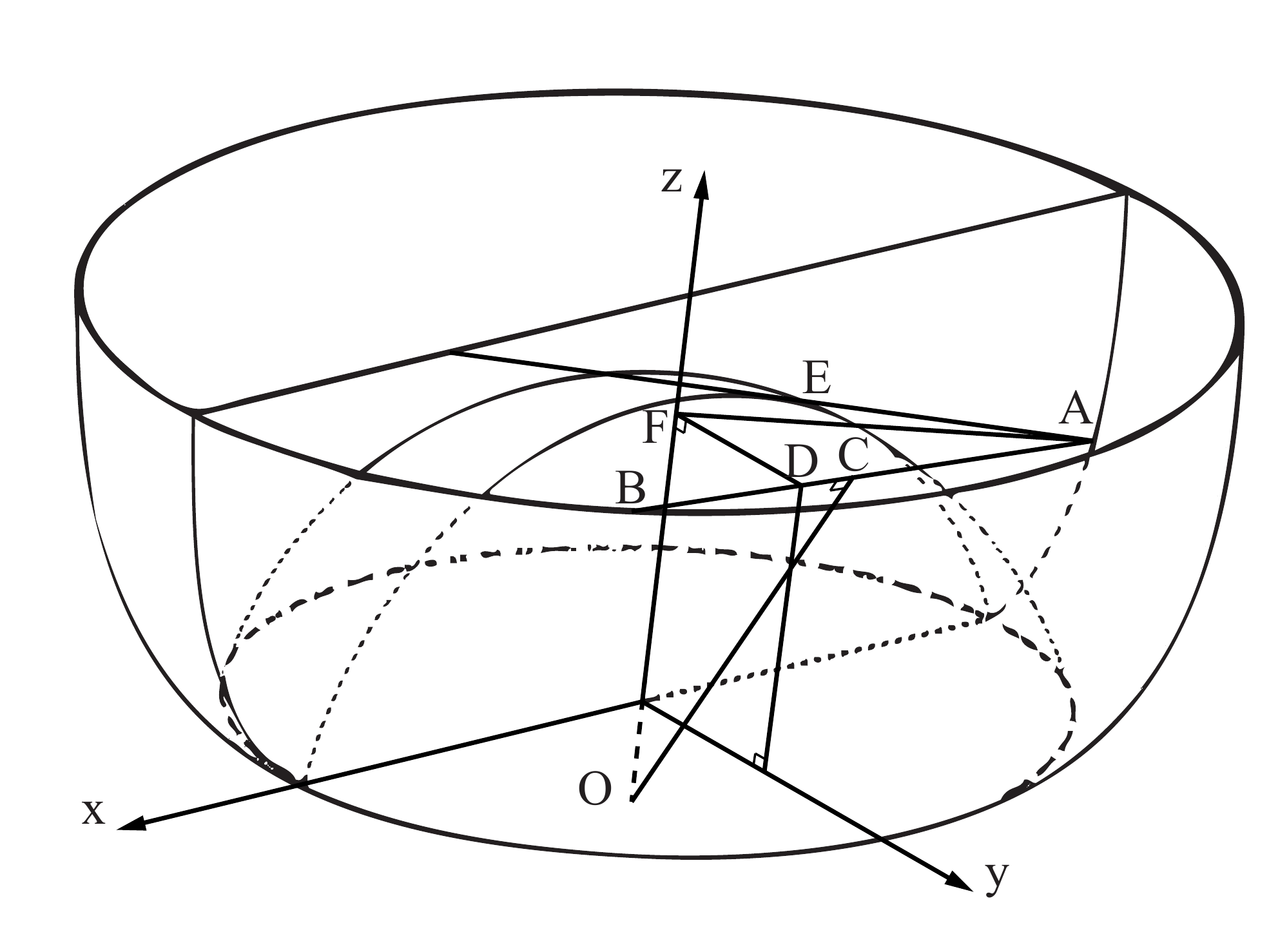}
\caption{Illustration of the proposed 3D modification of the String method. O is the center of the sphere of the photo cathode. AB is a new critical ray, which is tangential to the photocathode surface at point C, and intersects with the entry aperture at point B.
AB's projection on the x-z plane is AF.
AE is the critical ray of the String method, and
AF has a larger incident angle than AE.
The 3D modification uses the new critical ray AB as the key input.
}
\label{fig:modify}
\end{figure}

\begin{figure}[h]
\includegraphics[width=7.5cm]{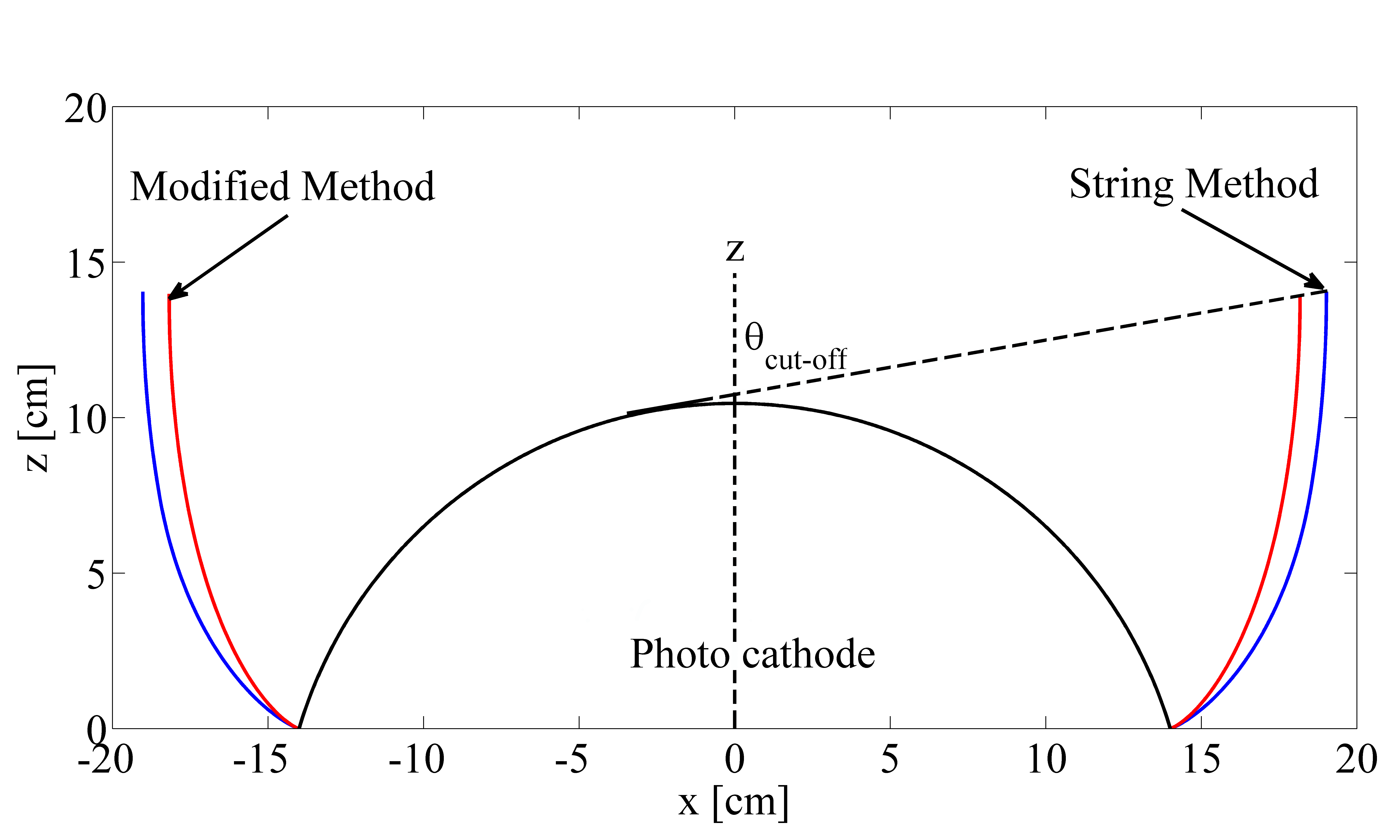}
\caption{2D profile of the String method and the modified method for 80$^\circ$ cut-off angle.
}
\label{fig:profiles}
\end{figure}

The 3D modified design also determines the profile curve first, and then rotates the curve about the z axis to obtain the reflector surface.
The modification of the profile curve requires several iterations.
The first iteration uses the reflector surface given by the String method as input.
For later iterations, the reflector surface from the previous step is used as input.
After each iteration, the concentrator becomes smaller, and the collection efficiency is increased.
The process stops when no obvious change of the size of the entry aperture is found.

Similar to that in the String method, each iteration starts with a point on the PMT aperture,
after which the relevant critical ray is chosen and the profile curve grows by an infinitesimal amount according to the slope determined by the critical ray.
The only difference between the original and modified methods is the way the critical ray is selected: the
modified method considers the 3D surface of the photo cathode.
Based on Fig.~\ref{fig:modify}, the critical ray, line AB,  for point A must be tangential to the 3D surface of the photo cathode
and intersect with the entry aperture of the concentrator.
The projection of AB on the x-z plane, AF, has a larger incident angle $\theta$ than that of AE, which is the critical ray in the String method and is only tangential in the x-z plane. The slope of the concentrator is then updated according to AF.
As an example, for a cut-off angle of 80$^\circ$, the comparison of the 2D profile of
the String method and the modified method is shown in Fig.~\ref{fig:profiles}. The modified method has shrinked the opening.

All of the light rays incident upon the entry aperture surface with the same $\theta$ and $\phi$ are traced in Fig.~\ref{fig:NoReflection}.
The modification of a smaller entry aperture indeed offers a better solution for some incidental rays.
In particular, the modification is very effective when the shape of a photo cathode is significantly different from that of a plane.
For concentrators with a small $\theta_{cut-off}$, the modification is not significant.
More numerical results can be seen in Section~\ref{sec:performance}.

\subsection{Hexagonal opening concentrator}
\begin{figure}[h]
\centering
\subfigure[Circular opening reflectors]{
  \label{closepack:a}
  \includegraphics[width=0.3\textwidth]{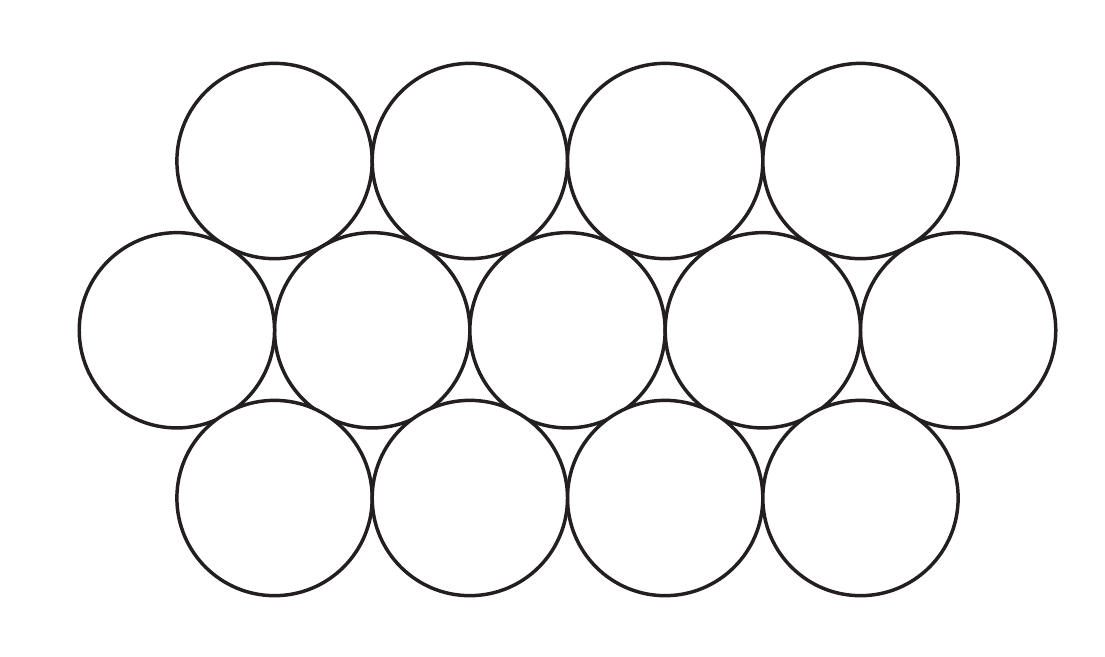}
  }
\subfigure[Hexagon opening reflectors]{
  \label{closepack:b}
  \includegraphics[width=0.3\textwidth]{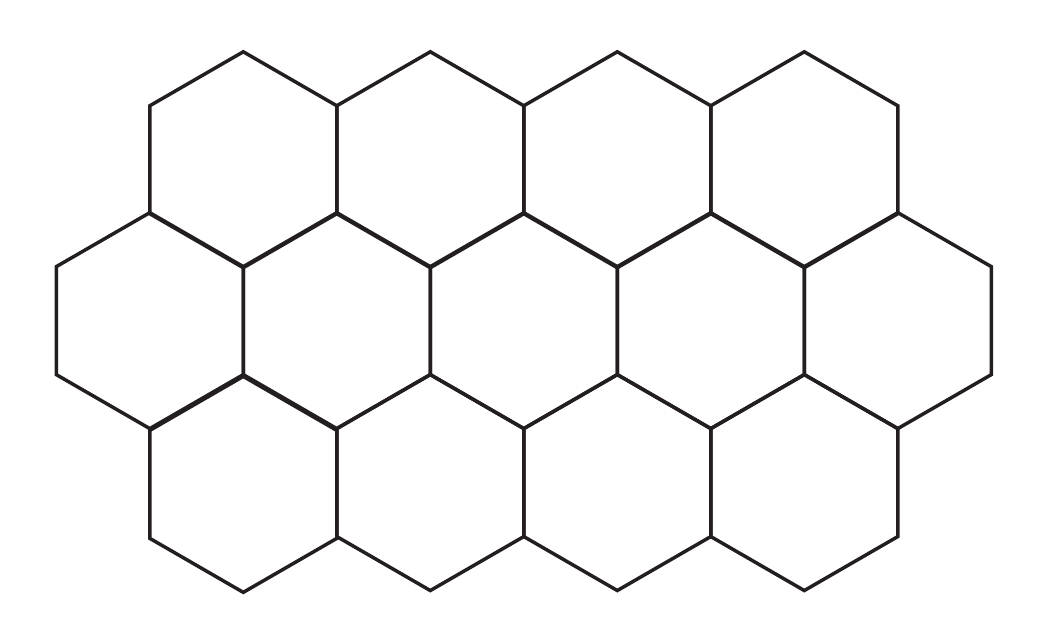}
  }
\caption{Illustrations of gapless packed concentrators with different opening shapes. (a) Circular opening concentrators cannot realize $100\%$ coverage because of the gap. (b) Hexagonal opening concentrators allow a coverage rate of $100\%$.}
\label{fig:closepack}
\end{figure}

The second consideration is to replace the circular opening of the concentrator with a hexagonal one~\cite{SPIE, ASP, POS}.
The motivation here is to increase the upper limit of $coverage$ to $100\%$, as shown is in Fig.~\ref{fig:closepack}.
With a circular opening, the upper limit of the photo cathode $coverage$ is 91\%.
Large-diameter PMTs, especially those used in neutrino experiments, are under water or similar liquid. Spherical head PMTs performs the best in resisting high pressure. Any change in shape is better made on the concentrator.

\begin{figure}[h]
\includegraphics[width=7.5cm]{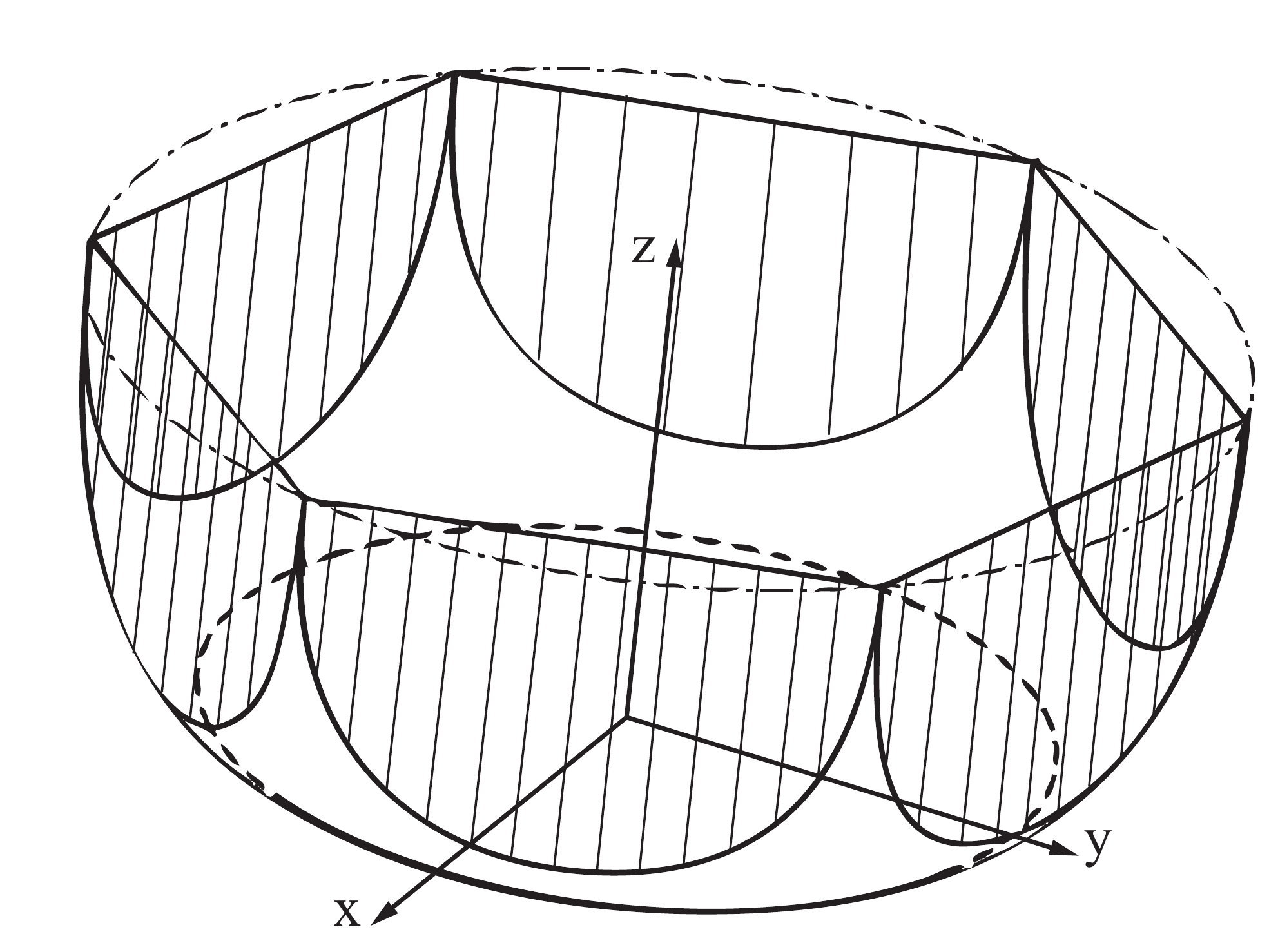}
\caption{Replacement of a circular opening concentrator with a hexagonal opening one. The hexagonal opening concentrator is achieved by using the inscribed hexagon of the entry aperture to cut along the z axis.}
\label{fig:hex}
\end{figure}

The hexagonal opening concentrator is constructed in the following way. An inscribed hexagon is first generated at the entry aperture, and then all of the sections outside of the hexagon are cut off along the z direction as shown in Fig.~\ref{fig:hex}.
Planes along the z axis can maximally maintain the downward trend of incoming light.

\section{Performance and cost}
\label{sec:performance}

\subsection{Performance comparison}
\begin{figure}[h]
\includegraphics[width=8.5cm]{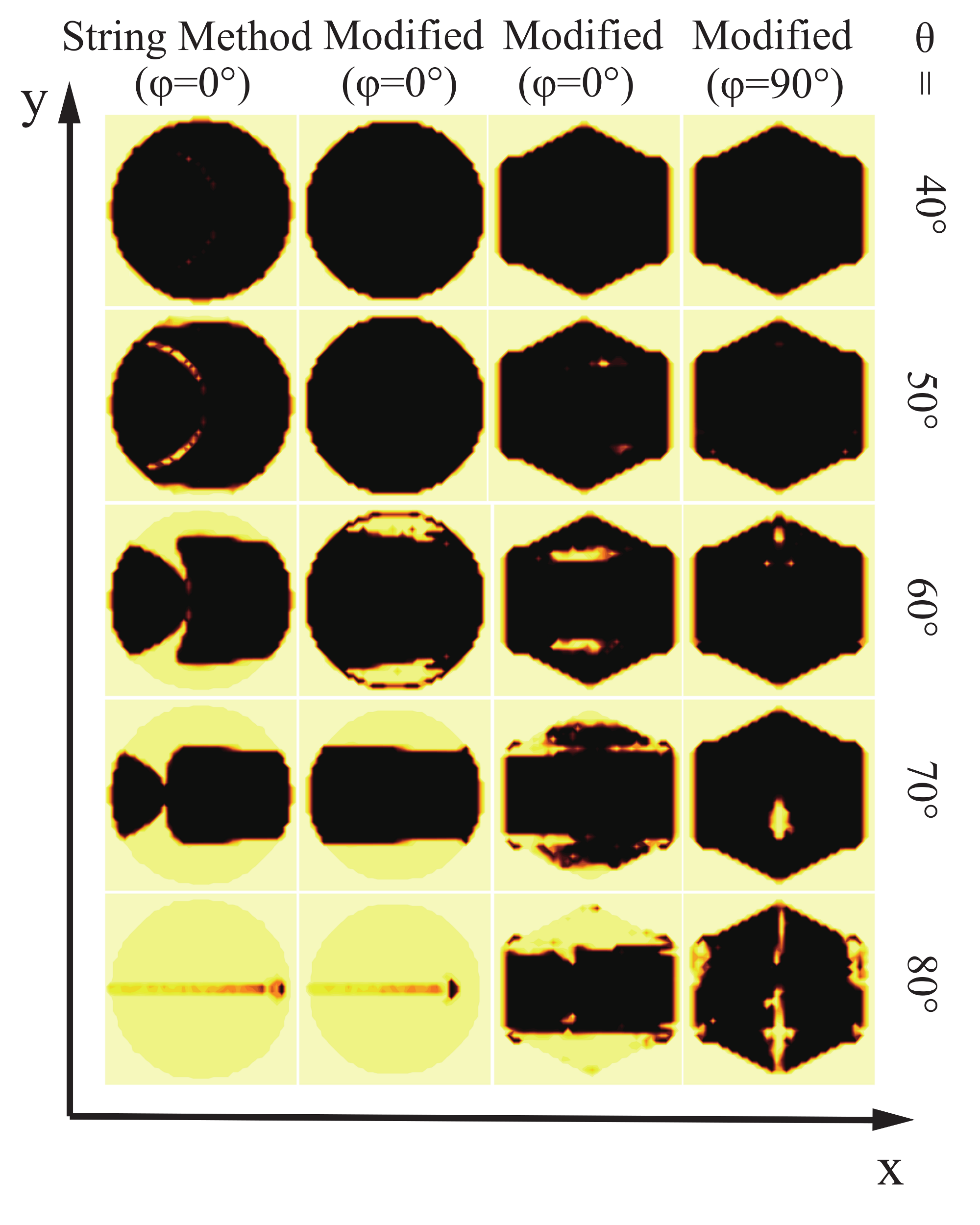}
\caption{Distribution of efficiency over incident position for concentrators designed using different methods. The definitions of $\theta$ and $\phi$ are identical to those in Fig.~\ref{fig:intro}.
The dark areas represent rays
entering the concentrators and hit the photocathode, while the light areas represent rays that escape from the entry aperture.}
\label{fig:2d-analysis}
\end{figure}

Taking a cut-off angle of 80$^\circ$ as an example, the light concentrator performance with the String method, the modified String method, and
the modified String method with hexagonal opening is compared in Fig.~\ref{fig:2d-analysis}.
Light rays were scanned with incidental angles $\theta$ from 40$^\circ$ to 80$^\circ$ and were shot uniformly on the entry aperture.
For the hexagonal opening, two symmetrical axes were tested.
With a small incident angle 40$^\circ$ or 50$^\circ$, nearly all of the rays were collected and the difference between these designs is not significant.
With a larger incident angle 60$^\circ$ or 70$^\circ$,
light rays from the arc region near the edge of the entry aperture and the crescent region near the center began to miss the photo cathode in the String method cone. In the modified String method cone with a circular opening, the escaped light rays were only from regions far from the center. Such a difference can also be seen in Fig.~\ref{fig:NoReflection}.
For the hexagonal opening, although the initial goal is to achieve a maximal coverage of concentrators, it also cuts off some of inefficiency regions and can further enhance the geometrical collection efficiency.
At 80$^\circ$, i.e. incident angle equals cut-off angle, incoming light rays have no chance to be detected except that they are in the profile plane, but the hexagonal opening changed their property and most of them were saved.
The modified String method with a hexagonal opening has most of its area in black, i.e. positive detection, and
it has the highest detection efficiency over the entry aperture.
A minor defect was found in our implementation, for example in the center area of the bottom right sub-figure of Fig.~\ref{fig:2d-analysis}, where no efficiency loss is expected. It happens in creating the triangular facets to construct the Geant4 models,
since a small discrepancy exists between the theoretical shape and the triangular facets approximation due to the limitation of software calculations.

\subsection{PMT number and efficiency scan}
To facilitate the design of a neutrino detector, the key parameters, $N_{PMT}$ and $\varepsilon$, are scanned with different cut-off angle requirements and for different light concentrator designs, namely, (1) the String method, (2) the modified method with a circular opening, (3) the String method with a hexagonal opening, and (4) the modified method with a hexagonal opening.

\begin{figure}[!h]
\centering
\includegraphics[width=8cm]{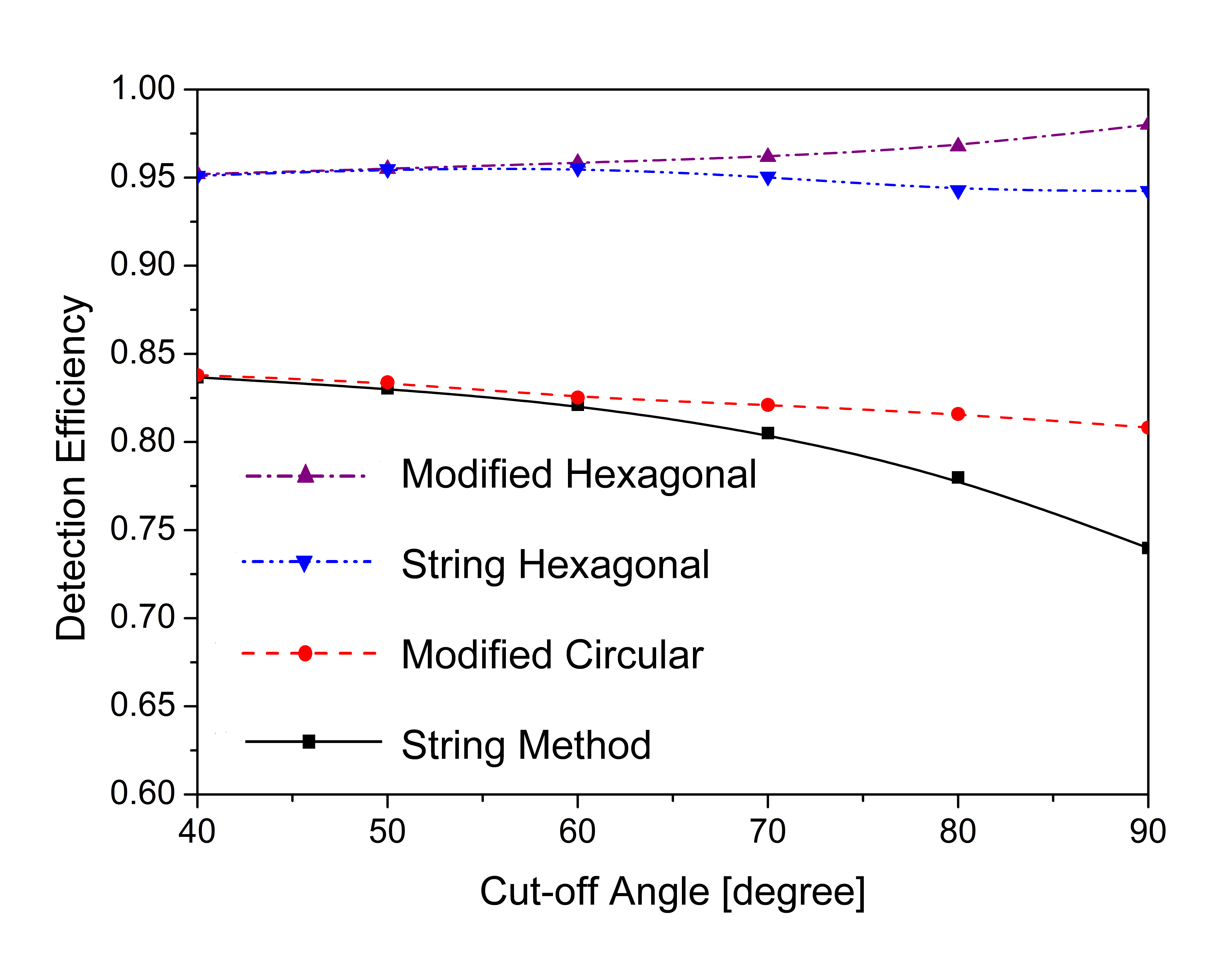}
\caption{Detection efficiency $\varepsilon$ versus cut-off angle for four designs.}
\label{fig:EffScan}
\end{figure}

\begin{figure}[!h]
\centering
\includegraphics[width=8cm]{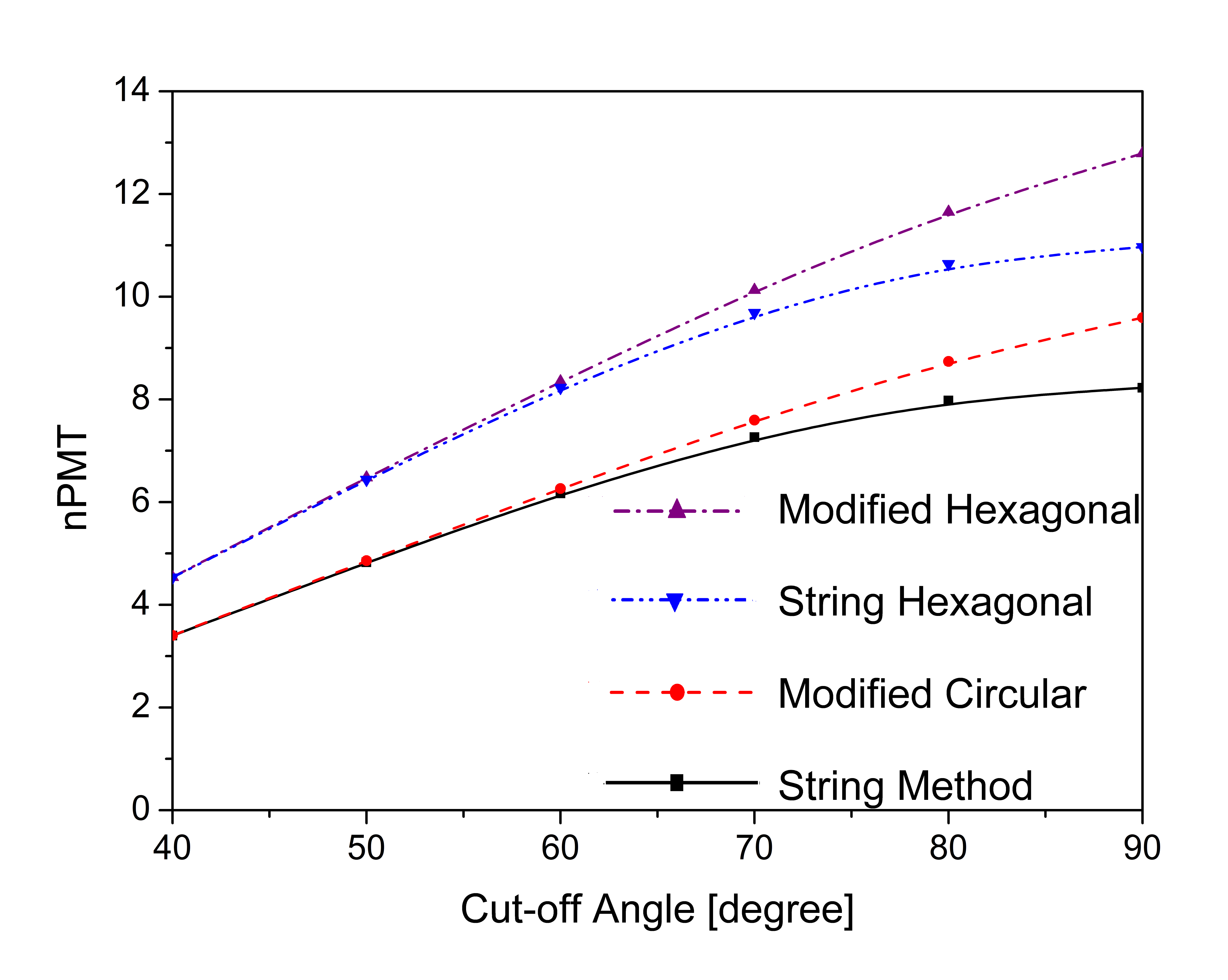}
\caption{Number of PMTs per unit area (one square meter) versus cut-off angle for four types of designs.}
\label{fig:NpmtScan}
\end{figure}

For $\theta_{cut-off}=80^\circ$, Table~\ref{tab:per} tabulates the detection efficiency $\varepsilon$, $coverage$, $\varepsilon_{col}$, and $N_{PMT}$, where $N_{PMT}$ is the maximum number of PMTs per unit area (one square meter).
In Figs.~\ref{fig:EffScan} and~\ref{fig:NpmtScan}, $\varepsilon$ and $N_{PMT}$ are plotted versus the cut-off angle.
Based on the table and figures,
by employing concentrators designed with the modified String method and hexagonal opening it is possible to reach a maximal photocathode coverage and a geometrical collection efficiency of 98\%, and it also saves about 20\% of PMTs than without concentrators.

\begin{table}[!h]
\begin{center}
\caption{Comparison of different concentrators with $\theta_{cut-off}=80^\circ$. Parameters listed include $coverage$, collection efficiency  $\varepsilon_{col}$,
total efficiency $\varepsilon$, and the number of PMTs per unit area $nPMT(m^{-2})$.
The factors are calculated assuming that the incident position and direction of scintillation light are uniformly distributed on the entry aperture.}
\label{tab:per}
 \begin{tabular}{c c c c c c}
 \hline
               & $coverage$ (\%)   &  $\varepsilon_{col}$ (\%) & $\varepsilon$ (\%) &$nPMT(m^{-2})$  \\ \hline
 No reflectors      &91  &100 &91 &14.73 \\ \hline
 String method      &91  &86  &78 &7.97\\ \hline
 Modified circular  &91  &90  &82 &8.73\\ \hline
 String hexagon     &100 &94  &94 &10.64\\ \hline
 Modified hexagon   &100 &97  &97 &11.65\\ \hline
 \end{tabular}
\end{center}
\end{table}


\subsection{Timing feature}
\begin{figure}[h]
\includegraphics[width=8cm]{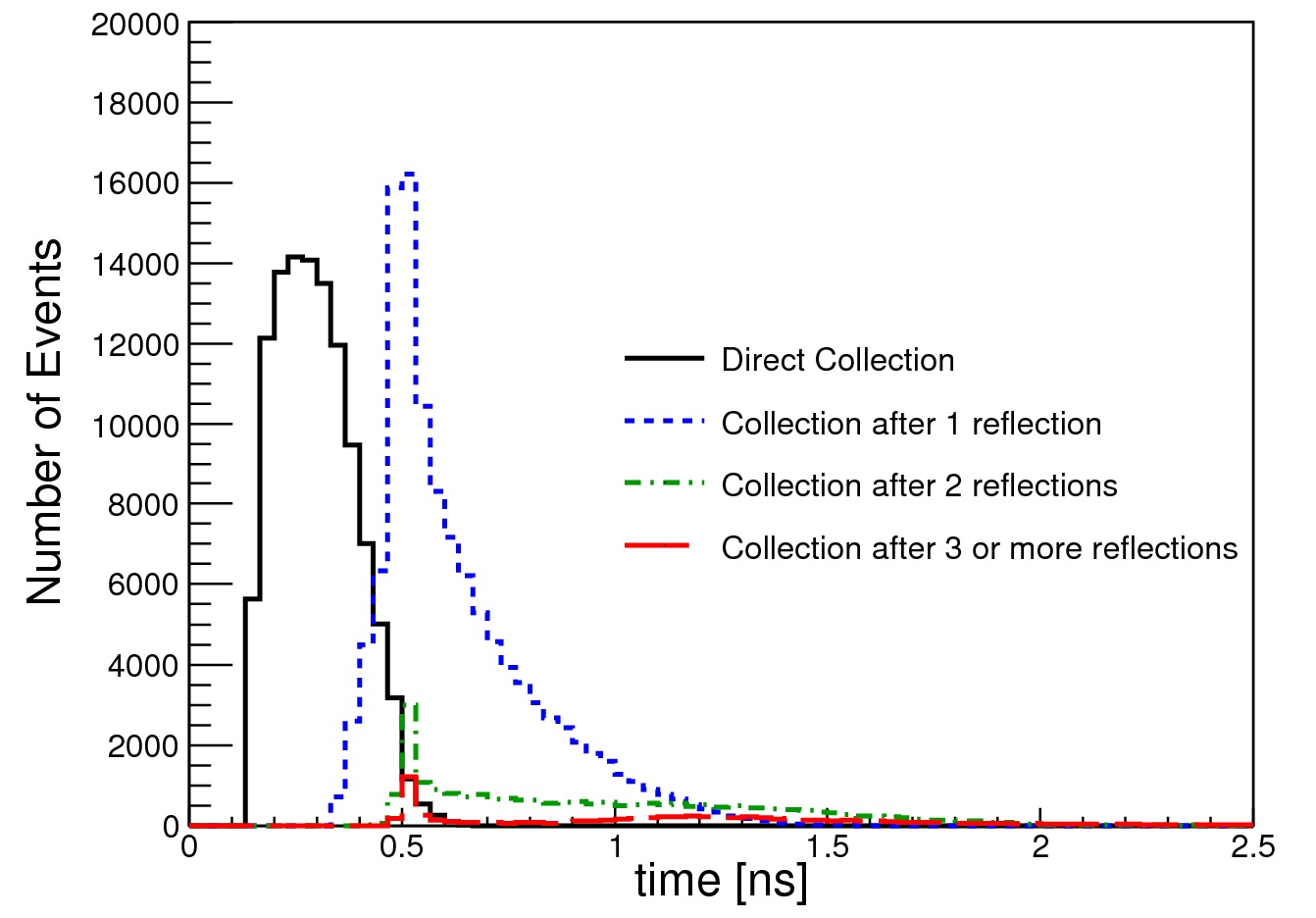}
\caption{Distribution of the time of incident rays from passing through the entry aperture to being collected by the photocathode. The first peak represent direct detection and the second peak for one reflection.}
\label{fig:Time}
\end{figure}

Besides the collection performance of the concentrator, the extra time delay of signals resulting from reflections may be relevant for some measurements. According to the simulation result for the light concentrator designed using the modified String method with a hexagonal entry aperture and $\theta_{cut-off}=80^\circ$, the time incident rays take from passing through the entry aperture to being collected by the photocathode has a spread (RMS) of less than 0.5 ns, as shown in Fig.~\ref{fig:Time},
and the first peak represents the direct detection and the second peak indicates one reflection.

\section{Conclusion}
\label{sec:conclusion}
The features of light concentrators designed by the String method are studied. Considering the 3D profile of the PMT,
the String method is modified to improve the collection efficiency of photons.
A hexagonal aperture is considered to realize a gapless arrangement and reach an upper limit of 100\% coverage.
Concentrators with a wide field view of 90$^\circ$ and a high efficiency of above 98\% are attained with these improvement.
The requested number of PMT is higher than the string method, but smaller than without any concentrator.
These improvements are important for an experiment to achieve the maximum light yield and the best energy resolution.
The key parameters of detector design, $N_{PMT}$ and $\varepsilon$, were provided for a wide range of cut-off angles and they can be directly used for future experiment design. Materials for the reflector will be investigated next and more detailed simulation studies including refraction on PMT glass interface will be carried out.

\section*{Acknowledgement}
This work is supported in part by the National Natural Science Foundation of China (Nos.~11235006 and 11475093), the Key Laboratory of Particle \& Radiation Imaging (Tsinghua University), and the CAS Center for Excellence in Particle Physics (CCEPP).


\begin{thebibliography}{99}

\bibitem{Welford} W.T. Welford, R. Winston, The Optics of Nonimaging Concentrators, Academic Press,Inc., New York, 1978.
\bibitem{Guan:2006gsa} R. Winston, J. Minano and P. Benitez, Nonimaging Optics, Academic Press, Inc., New York, 2004: 89-92.

\bibitem{SNO}        G. Doucas, et al., Nucl. Instrum. Methods Phys. Res. A 370 (1996) 579.
\bibitem{SNOThesis}  Guy Ren$\acute{e}$ Ouellette (Master thesis), University of British Columbia, 1991.
\bibitem{BX}         L. Oberauer, C. Grieb, F. von Feilitzsch, and I. Manno, Nucl. Instrum. Methods Phys. Res. A 530 (2004) 453.

\bibitem{SPIE}       F.~H$\acute{e}$nault, et al., Proc. of SPIE Vol. 8834, 883405 (2013).
\bibitem{ASP}        A.~Okumura, Astroparticle Physics 38, 18 (2012).
\bibitem{POS}        S.~Querchfeld, J.~Rautenberg and K.~Kampert, PoS (ICRC2015) 677 (2015).

\bibitem{Minfang}    M.~Yeh, et al., Nucl. Instrum. Methods Phys. Res. A 660 (2011) 51.
\bibitem{LAB}        M.~Li, et al., Nucl. Instrum. Methods Phys. Res. A 830 (2016) 303.
\bibitem{GuoZY}      Z.~Guo, et al., arXiv:1708.07781 (2017).
\bibitem{CHESS}      J. Caravaca, et al., Phys. Rev. C 95 (2017) 055801.
\bibitem{Hanyu}      H.~Wei, Z.~Wang, S.~Chen, Phys. Lett. B 769 (2017) 255.
\bibitem{GeoMK}      Z.~Wang and S.~Chen, arXiv:1709.03743 (2017).
\bibitem{Jinping}    J.F.~Beacom, et al., Chinese Physics C 41 (2017) 023002.
\bibitem{JPGeo}      L.~Wan, G.~Hussain, Z.~Wang, and S.~Chen, Phys. Rev. D 95 (2017) 053001.

\bibitem{LENA}       M.~Wurm, et al., Astropart. Phys. 35 (2012) 685.
\bibitem{JUNO}       F.~An, et al., J. Phys. G: Nucl. Part. Phys. 43 (2016) 030401.
\bibitem{JUNO2}      M.~B.~Avanzini, et al., arxiv:1612.05444 (2016).

\bibitem{g41} S.~Agostinelli {\it et al}., Nucl. Instrum. Methods Phys. Res. A 506 (2003) 250.
\bibitem{g42} J.~Allison {\it et al}., IEEE Trans. Nucl. Sci. 53 (2006) 270.

\bibitem{XP1807}     http://hzcphotonics.com/products/XP1807.pdf


\end{thebibliography}
\end{document}